\documentclass[epj]{webofc}
\pdfoutput=1
\usepackage[utf8]{inputenc}
\usepackage[varg]{txfonts}   % Web of Conferences font
\usepackage{booktabs}
\usepackage{xcolor}
\definecolor{darkred}{rgb}{0.4,0.0,0.0}
\definecolor{darkgreen}{rgb}{0.0,0.4,0.0}
\definecolor{darkblue}{rgb}{0.0,0.0,0.4}
\usepackage[bookmarks,linktocpage,colorlinks,
    linkcolor = darkred,
    urlcolor  = darkblue,
    citecolor = darkgreen]{hyperref}
\wocname{EPJ Web of Conferences}
\woctitle{Lattice2017}
\begin{document}
%%%%%%%%%%%%%%%%%%%%%%%%%%%%%%%%%%%%%%%%%%%%%%%%%%%%%%%%%%%%%%%%%%%%%%%%%%%%%
%
\selectlanguage{english}
%----------------------------------------------------------------------------
\title{%
$B \rightarrow \pi \ell \nu$ with M\"{o}bius Domain Wall Fermions
}
%----------------------------------------------------------------------------
\author{%
\firstname{Brian} \lastname{Colquhoun}\inst{1}\fnsep\thanks{Speaker, \email{brian.colquhoun@kek.jp}} \and
\firstname{Shoji} \lastname{Hashimoto}\inst{1,2} \and
\firstname{Takashi}  \lastname{Kaneko}\inst{1}
% etc.
}
%----------------------------------------------------------------------------
\institute{%
High Energy Accelerator Research Organisation (KEK), Tsukuba 305-0801, Japan
\and
School of High Energy Accelerator Science, SOKENDAI (The Graduate University for Advanced Studies), Tsukuba 305-0801, Japan
}
%----------------------------------------------------------------------------
\abstract{%
We report on the status of our calculation of the exclusive semileptonic decay, $B\rightarrow \pi \ell \nu$; a key process in the determination of the CKM matrix element $|V_{ub}|$. The M\"{o}bius domain wall action is used for both light and heavy quarks on gauge ensembles that include the effects of $2+1$ flavours of quarks in the sea at three values of the lattice spacing: $a\approx 0.08~\mathrm{fm}$, $a\approx0.055~\mathrm{fm}$, and $a\approx0.044~\mathrm{fm}$. Pion masses go down to $300~\mathrm{MeV}$ while heavy quarks masses are as large as $2.44m_c$. We present preliminary results of form factors from this process, showing dependence on momentum transfer, lattice spacing, and the heavy quark mass.
}
%----------------------------------------------------------------------------
\maketitle
%----------------------------------------------------------------------------
\section{Introduction}\label{intro}

The semileptonic process $B\rightarrow \pi \ell \nu$ is important in the determination of the element $|V_{ub}|$ of the Cabibbo-Kobayashi-Maskawa matrix, which describes the mixing between quark flavours. We report on the progress of our lattice QCD calculation of this process.

Effective theory approaches to $b$ quarks are often used on the lattice to avoid large discretisation errors arising from the large quark mass. It is, however, possible to simulate $b$ quarks by carrying out calculations at multiple smaller quark masses where discretisation effects are under control and extrapolating to the physical mass. We carry out this procedure for heavy quark masses $m_h \geq m_c$ using M\"{o}bius Domain Wall Fermions. Doing so means being able to use the same action for light and heavy quarks, avoiding any difficulties that can arise from using a mixed action for the valence quarks.

\section{Form Factors}\label{sec-1}
The CKM matrix element $|V_{ub}|$ relates to the differential decay rate of the process $B\rightarrow \pi \ell \nu$, measured at BaBar and Belle~\cite{delAmoSanchez:2010af,Ha:2010rf,Lees:2012vv,Sibidanov:2013rkk}, by,
\begin{equation}\label{eq:ckm_extraction}
\frac{\mathrm{d}\Gamma\left(B\rightarrow\pi\ell\nu\right)}{\mathrm{d}q^2}=\frac{G^2_F\left| V_{ub}\right|^2}{24\pi^3}\left|p_\pi\right|^3 \left|{f_+} (q^2)\right|^2.
\end{equation}
The form factor $|f_+(q^2)|$ needs to be theoretically calculated to extract $|V_{ub}|$.

The semileptonic decay of a $B$ meson to a $\pi$ can be described by the matrix elements,
\begin{equation}
\langle{\pi(k_\pi)} |V^{\mu}|{B(p_B)}\rangle =
f_{+}(q^2) \left[p_B^{\mu} + k^{\mu}_\pi - \frac{m_B^2 - m_\pi^2}{q^2} q^{\mu} \right] +
f_{0} (q^2) \, \frac{m^2_B - m^2_\pi}{q^2} \, q^{\mu} \, ,
\end{equation}
where $f_{+}(q^2)$ and $f_{0}(q^2)$ are the vector and scalar form factors from this process, $p_B$ and $k_\pi$ are the 4-momenta of the $B$ and $\pi$ respectively, and $m_B$ and $m_\pi$ are their masses. The momentum transfer is $q^\mu=p^\mu_B-k^\mu_\pi$. At $q^2=0$ there exists a constraint, $f_+(0)=f_0(0)$.

A convenient alternative parametrisation when working with heavy quarks is in terms of the heavy quark velocity, $v^\mu=p^\mu_B/m_B$, such that~\cite{Burdman:1993es},
\begin{equation}
  {\langle \pi(k_\pi) | V^{\mu} | B(v) \rangle =2\left[f_1\left(v\cdot k_\pi\right)v^\mu +f_2\left(v \cdot k_\pi\right)\frac{k^\mu}{v \cdot k_\pi}\right]},
\end{equation}
where $v\cdot k_\pi$ is the pion energy measured in the rest frame of the $B$ meson:
\begin{equation}
v\cdot k_\pi \equiv E_\pi = \frac{m^2_B+m^2_\pi-q^2}{2m_B}.
\end{equation}
The form factors then relate to the vector matrix elements between the $B$ and the $\pi$ through,
\begin{align}
  f_1\left(v\cdot k_\pi\right)+f_2\left(v \cdot k_\pi\right)&=\frac{\langle \pi(k_\pi) | V^{0} | B(v) \rangle}{2};\\
  f_2\left(v \cdot k_\pi\right)&=\frac{\langle \pi(k_\pi) | V^{i} | B(v) \rangle}{2}\frac{v \cdot k_\pi}{k^i_\pi}.
\end{align}
In the limit $v \cdot k_\pi \rightarrow 0$, there is a prediction of a pole dominance,
\begin{equation}\label{eq:pole_dominance}
  \lim_{v \cdot k_\pi \to 0}f_2\left(v \cdot k_\pi\right)=g\frac{f_{B^*}\sqrt{m_{B^*}}}{2f_\pi}\frac{v \cdot k_\pi}{v \cdot k_\pi + \Delta_B},
\end{equation}
with $m_{B^*}$ the mass of the vector meson $B^*$, $\Delta_B$ the hyperfine splitting, $f_{B^*}$ and $f_\pi$ the $B^*$ and $\pi$ decay constants respectively, and $g$ the $B^*B\pi$ coupling.

This parametrisation relates back to the $f_+(q^2)$ and $f_0(q^2)$ form factors relevant for comparison with experiment through,
\begin{align}
\label{eq:form_factor_transform}
  f_+\left(q^2\right)=&\sqrt{m_B}\left[\frac{f_2\left(v \cdot k_\pi\right)}{v\cdot k_\pi}+\frac{f_1\left(v\cdot k_\pi\right)}{m_B}\right];\\
  f_0\left(q^2\right)=&\frac{2}{\sqrt{m_B}}\frac{m^2_B}{m^2_B-m^2_\pi}\left[f_1\left(v\cdot k_\pi\right)+f_2\left(v \cdot k_\pi\right)\right.\nonumber \\
    &\left.-\frac{v\cdot k_\pi}{m_B}\left(f_1\left(v\cdot k_\pi\right)+\frac{m^2_\pi}{\left(v\cdot k_\pi\right)^2}f_2\left(v \cdot k_\pi\right) \right)\right].
\end{align}
In the heavy quark limit there then appears a straightforward scaling between these parametrisations:
\begin{align}
  f_+\left(q^2\right)\sim&\sqrt{m_B};\\
  f_0\left(q^2\right)\sim&\frac{1}{\sqrt{m_B}}.
\end{align}

\section{Lattice Calculation}
In our calculation, we use ensembles that include the effects of $2+1$ flavours of M\"{o}bius Domain Wall Fermions~\cite{Brower:2012vk} in the sea. We also use the M\"{o}bius Domain Wall Fermion action for the valence quarks.

Due to the increasing discretisation errors with increasing mass, we do not reach the physical $b$ quark mass. Instead, we use multiple values of a heavy quark mass, $m_h$, from the charm mass, $m_c$, up to $2.44\times m_c$. For comparison, the ratio $m_b/m_c$ has been measured on the lattice to be very close to 4.5~\cite{Maezawa:2016vgv,Bussone:2016iua,Chakraborty:2014aca}. From our values of $m_h$ we can perform an extrapolation towards $m_b$. We use a subset of the heavy quark masses used in the calculation of decay constants~\cite{Fahy:2017enl}. This work is also effectively an extension of the $D\rightarrow \pi\ell\nu$ calculation, using the $D$ meson as the lightest of the heavy-light mesons~\cite{Kaneko:2017sct,kaneko:lat17}.

We use three values of the lattice spacing: $a\approx 0.08~\mathrm{fm}$, $a\approx0.055~\mathrm{fm}$, and $a\approx0.044~\mathrm{fm}$, corresponding to cutoffs $a^{-1}\approx2.45~\mathrm{GeV}$, $a^{-1}\approx3.61~\mathrm{GeV}$ and $a^{-1}\approx4.50~\mathrm{GeV}$. These ensembles have dimensions $32^3\times 64$, $48^3\times 96$ and $64^3\times 128$ respectively. The pion masses used range from $m_\pi\approx 500~\mathrm{MeV}$ down to $m_\pi\approx 300~\mathrm{MeV}$.

We work only in the rest frame of the $B$ meson, and therefore to get a range of transfer momenta we simulate the pions at three-momentum $p=(n_x,n_y,n_z)\times 2\pi/L$, where $L$ is the spatial length in lattice units and $n_i\in \{-1,0,1\}$ for $i=x,y,z$. All permutations of a given momentum $p$ are calculated and averaged before fitting to improve statistics.

To improve our signal, we apply a spatial smearing function to each end of our correlators. We also generate two-point correlators with local sinks, allowing for the determination of, for example, the decay constants.

The numerical calculation of the correlators was performed using the IroIro++ code~\cite{Cossu:2013ola}.

\section{Results}
\subsection{Correlator Fits}\label{subsec:corr_fits}

We generate two sets of correlators: two-point correlators for the pseudoscalar mesons $B$ and $\pi$, and three-point correlators with the mesons at the source and sink, with vector operators inserted at each intermediate time $t$. We set the time separation, $T$, between the source and the sink such that it is long enough to extract a plateau when fitting. We choose $T=28$ on the coarsest lattice, which has volume $32^3\times 64$, and scale proportionally on the larger lattices so that the physical separation is kept constant.

Properly normalised, the three-point correlators can be described by,
\begin{equation}
C(t)= \frac{\langle{0}|P|\pi(k)\rangle \langle \pi(k) | V| B(p) \rangle \langle B(p) | P | 0 \rangle }{2E_\pi E_B}\left(\exp\left(-E_\pi t\right)+\exp\left(-E_{B}\left(T-t\right)\right)\right).
\end{equation}
To isolate the matrix elements $\langle \pi(k) | V| B(p) \rangle$ that relate to the form factors we want, we do a simultaneous fit of the three-point correlators with two-point correlators at sufficiently large time separations $t$ and $T-t$.

\begin{figure}[!bt]
  \centerline{
    \includegraphics[width=0.85\textwidth]{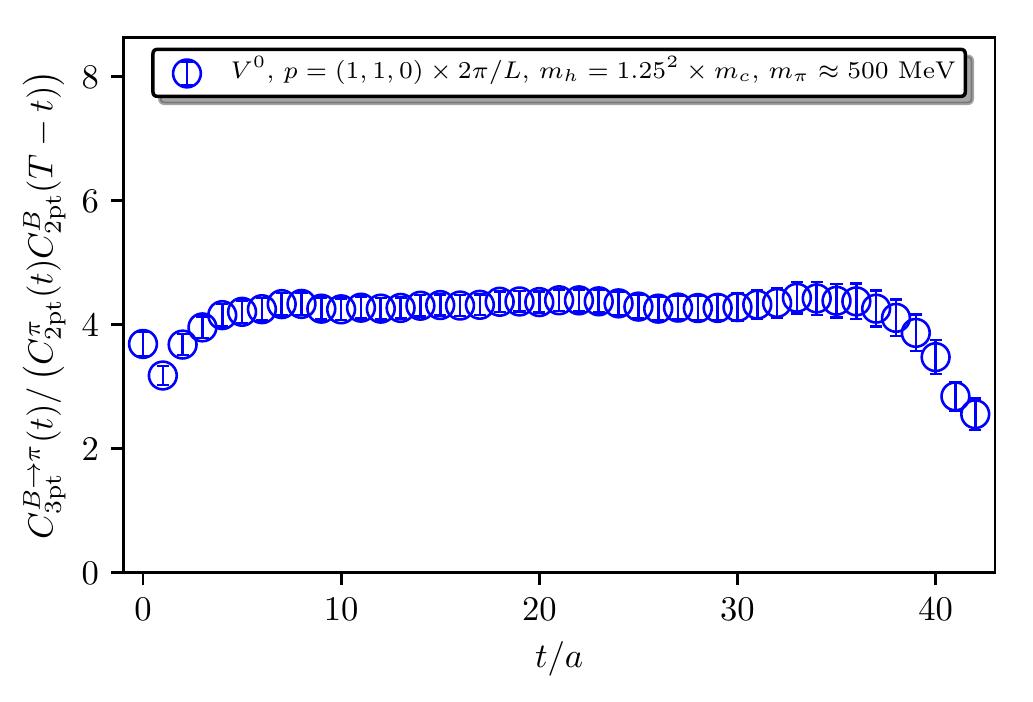}
    }
\caption{Plot of the ratio of the three-point correlator to the two-point correlators on the ensemble with $a^{-1}=3.61~\mathrm{GeV}$, with a pion mass $m_\pi\approx 500~\mathrm{MeV}$ and momentum $p=(1,1,0)\times 2\pi/L$. The heavy quark mass in this example is $m_h=1.25^2\times m_c$. The sink is at $T=42$.}
\label{fig:3pt_correlator}
\end{figure}

As a demonstration of the quality of our correlators, Figure~\ref{fig:3pt_correlator} shows the ratio of the three-point correlator to two-point correlators, $C^{B\rightarrow\pi}_{3\mathrm{pt}}/(C^{\pi}_{2\mathrm{pt}}(t)C^{B}_{2\mathrm{pt}}(T-t))$. In this case, we show the correlators resulting from the temporal vector operator on the ensemble with $a^{-1}\approx 3.61~\mathrm{GeV}$. The pion mass is $m_\pi\approx 500~\mathrm{MeV}$, the heavy quark mass is $m_h=1.25^2\times m_c$ and the momentum is the average of $p=(1,1,0)\times2\pi/L$ and all permutations. We observe a clear plateau stretching between $t=6$ and $t=36$.

We require a renormalisation constant $Z_{bl}$ for the $b$-to-light vector current. In principle, it is the same as that for the light-to-light renormalisation factor $Z_{ll}$ previously determined in~\cite{Tomii:2016xiv}. In order to eliminate possible discretisation effects for large $m_ha$, however, we adopt an improved estimate $Z_{bl}=\sqrt{Z_{bb}Z_{ll}}$ determined nonperturbatively by demanding the matrix element $\langle B | V | B \rangle=1$. The large effect on the wave function renormalisation observed in~\cite{Fahy:2017enl} can then be cancelled. For this we need an additional set of correlators. This overlaps with our work on inclusive $B$ meson decays presented at this conference where the same procedure is used~\cite{hashimoto:lat17}.

\subsection{Global Fit}
We perform global fits to the $f_1(v\cdot k_\pi)+f_2(v\cdot k_\pi)$ and $f_2(v\cdot k_\pi)$ form factor results from each of the ensembles to get results for physical parameters. For $f_1(v\cdot k_\pi)+f_2(v\cdot k_\pi)$ we use the fit form:
\begin{equation}\label{eq:f1_f2_fit}
f_1(v\cdot k_\pi)+f_2(v\cdot k_\pi)=C_0\left(1+C_{E_{\pi}}E_\pi+C_{E^2_{\pi}}E^2_\pi\right)\left(1+C_{a^2}a^2\right)\left(1+C_{m^2_{\pi}}m^2_\pi\right)\left(1+\frac{C_{m_h}}{m_h}\right),
\end{equation}
while for $f_2(v\cdot k_\pi)$, since we expect a contribution from the pole as in~(\ref{eq:pole_dominance}), we use,
\begin{equation}\label{eq:f2_fit}
f_2(v\cdot k_\pi)=\left[ D_0\left(1+D_{E_\pi}E_\pi\right)\left(1+D_{a^2}a^2\right)\left(1+D_{m^2_\pi} m^2_\pi\right)\left(1+\frac{D_{m_h}}{m_h}\right)\right]\frac{E_\pi}{E_\pi+\Delta_{B}}.
\end{equation}
We account for standard $a^2$ errors, pion mass dependence, dependence on the pion energy and dependence on the heavy quark mass. The form factor $f_2(v\cdot k_\pi)$ can be fitted with a term linear in $E_\pi$, but we include a quadratic term in~(\ref{eq:f1_f2_fit}) for the form factor $f_1(v \cdot k_\pi)+f_2(v \cdot k_\pi)$. The inclusion of higher order terms in $E_\pi$ in the fit returns coefficients consistent with zero, so we have chosen to exclude them here.

\subsection{Dependencies}
Here we demonstrate dependencies on various quantities in our calculation. We perform the global fits discussed above and set the values of the lattice spacing, pion mass and heavy quark mass appropriately. %The upper sets of lines/points correspond to the combination $f_1(v \cdot k_\pi)+f_2(v \cdot k_\pi)$, while the bottom set correspond to $f_2(v \cdot k_\pi)$ only. The effect of including a term for the pole dominance in $f_2(v\cdot k_\pi)$ is clear as the value rapidly goes to zero as $m_\pi\rightarrow 0$.

\begin{figure}[!bt]
  \centerline{
    \includegraphics[width=0.85\textwidth]{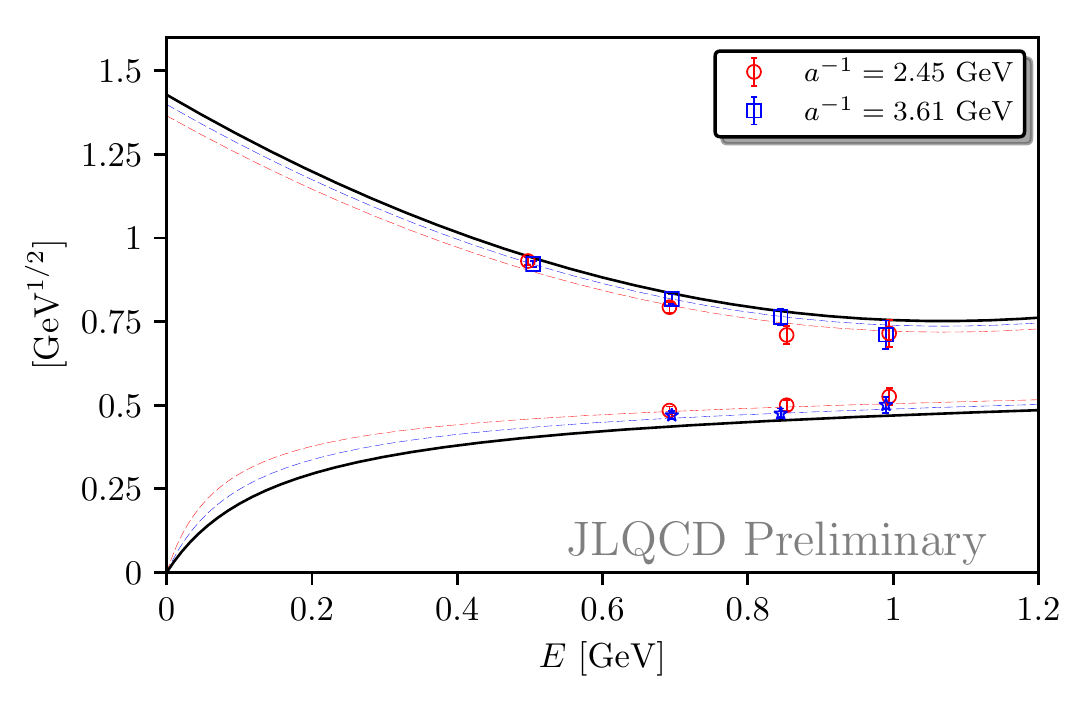}
    }
\caption{Dependence of the form factors on the lattice spacing. The upper set of lines and points correspond to the form factor combination $f_1(v \cdot k_\pi)+f_2(v \cdot k_\pi)$ and the lower to $f_2(v \cdot k_\pi)$. Using the global fit results, the dashed lines show the form factors with $m_\pi\approx 500~\mathrm{MeV}$ and $m_h=1.25^2\times m_c$ at $a^{-1}\approx 2.45~\mathrm{GeV}$ (red) and $a^{-1}\approx 3.61~\mathrm{GeV}$ (blue). The points show the data that were used in the global fit with those parameters: red circles for $a^{-1}\approx 2.45~\mathrm{GeV}$ and blue squares for $a^{-1}\approx 3.61~\mathrm{GeV}$. The solid black line gives the result in the $a^2\rightarrow 0$ limit.}
\label{fig:spacing_dependence}
\end{figure}

The dependence of the form factors on the lattice spacing is shown in Figure~\ref{fig:spacing_dependence}. In this case we take results with pion masses $m_\pi\approx 500~\mathrm{MeV}$ and use a heavy quark mass $m_h=1.25^2\times m_c$. The upper sets of lines/points correspond to the combination $f_1(v \cdot k_\pi)+f_2(v \cdot k_\pi)$, while the bottom set correspond to $f_2(v \cdot k_\pi)$ only. We plot dashed lines for the lattice spacing values: red for $a^{-1}\approx2.45~\mathrm{GeV}$ and blue for $a^{-1}\approx3.61~\mathrm{GeV}$. The solid black line indicates the result of the fit in the $a^2\rightarrow 0$ limit. The points show the individual form factor results with $m_\pi= 500~\mathrm{MeV}$: red circles for $a^{-1}\approx2.45~\mathrm{GeV}$ and blue squares for $a^{-1}\approx3.61~\mathrm{GeV}$. The lattice spacing dependence seen here is small.

\begin{figure}[!bt]
  \centerline{
    \includegraphics[width=0.85\textwidth]{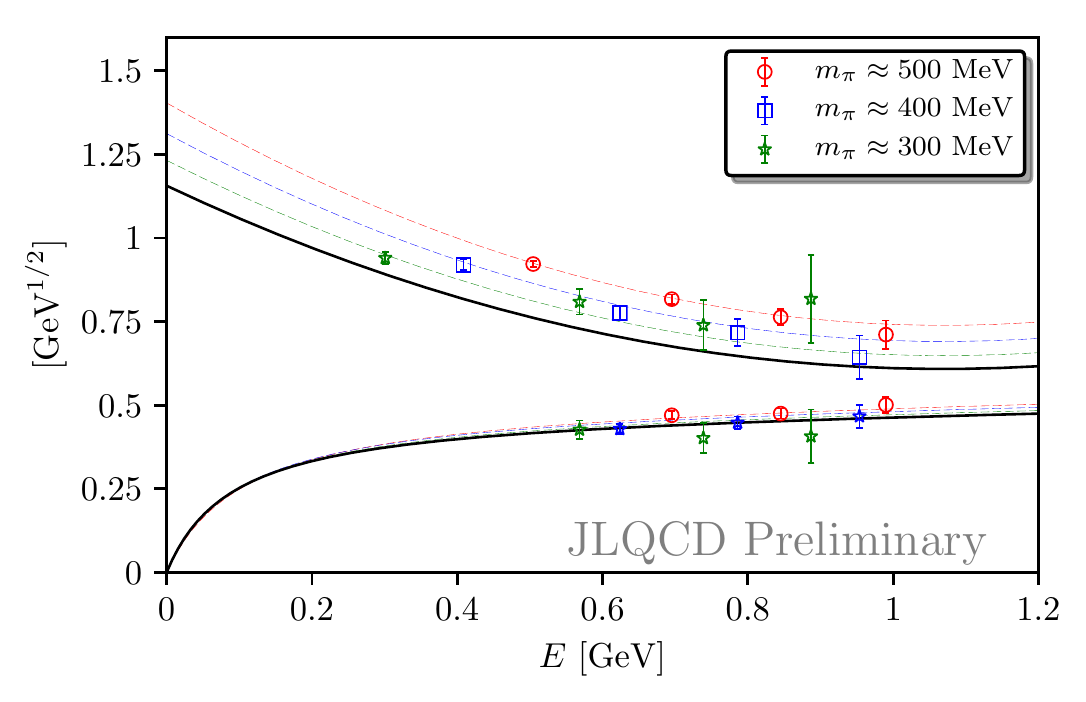}
  }
\caption{Dependence of the form factors on the pion mass. The upper set of lines and points correspond to the form factor combination $f_1(v \cdot k_\pi)+f_2(v \cdot k_\pi)$ and the lower to $f_2(v \cdot k_\pi)$. Using the global fit results, the dashed lines show the form factors on the ensembles with $a^{-1}\approx 3.61~\mathrm{GeV}$ using $m_h=1.25^2\times m_c$ and $m_\pi\approx 500~\mathrm{MeV}$ (red), $m_\pi\approx 400~\mathrm{MeV}$ (blue) and $m_\pi\approx 300~\mathrm{MeV}$ (green). The points show the data that were used in the global fit with those parameters: $m_\pi\approx 500~\mathrm{MeV}$ (red circles), $m_\pi\approx 400~\mathrm{MeV}$ (blue squares) and $m_\pi\approx 300~\mathrm{MeV}$ (green stars). The solid black line gives the result at $m_\pi=135~\mathrm{MeV}$.}
\label{fig:mpi_dependence}
\end{figure}

We similarly check the dependence on the pion mass. In the example given in Figure~\ref{fig:mpi_dependence} we show the results on the ensemble with $a^{-1}=3.61~\mathrm{GeV}$ and use $m_h=1.25^2\times m_c$ for the heavy quark. The dashed line (point) colours correspond to the various pion masses: red (circles) for $m_\pi\approx 500~\mathrm{MeV}$, blue (squares) for $m_\pi\approx 400~\mathrm{MeV}$ and green (stars) for $m_\pi\approx 300~\mathrm{MeV}$. The solid black line shows the result of an extrapolation to the physical value of the $\pi^0$, $m_{\pi^0}=135~\mathrm{MeV}$. In this case there is very little dependence on the pion mass for the form factor $f_2(v\cdot k_\pi)$. The $f_1(v \cdot k_\pi)+f_2(v \cdot k_\pi)$ form factor exhibits a larger dependence, but is still well under control.

\begin{figure}[!bt]
  \centerline{
    \includegraphics[width=0.85\textwidth]{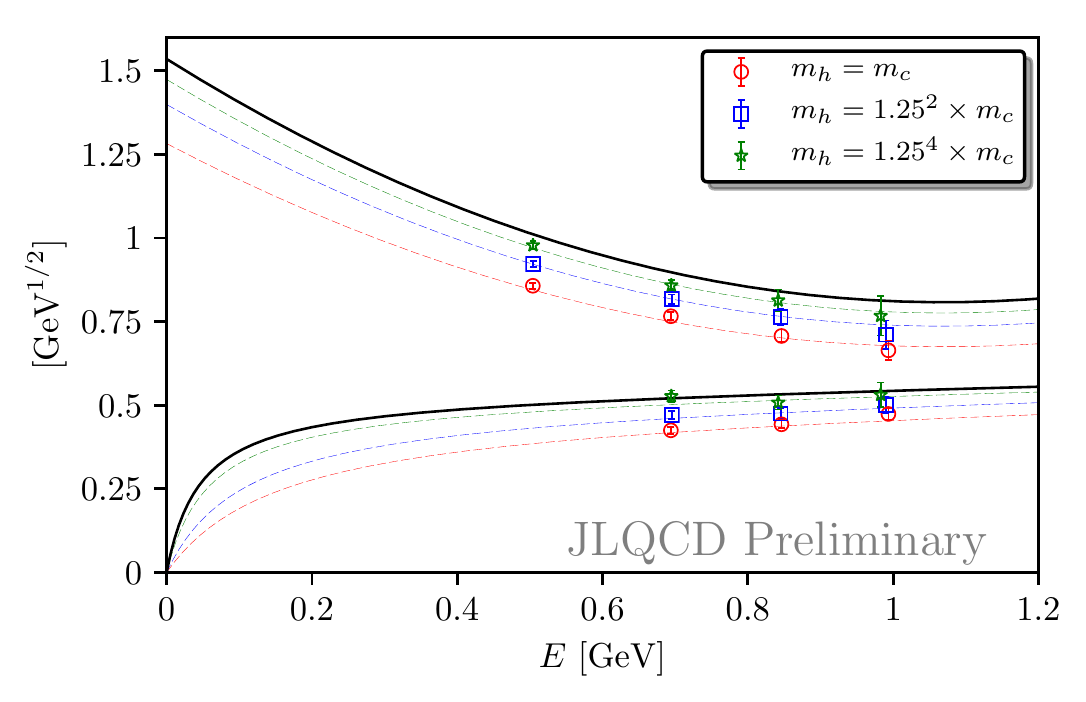}
    }
\caption{Dependence of the form factors on the heavy quark mass. The upper set of lines and points correspond to the form factor combination $f_1(v \cdot k_\pi)+f_2(v \cdot k_\pi)$ and the lower to $f_2(v \cdot k_\pi)$. Using the global fit results, the dashed lines show the form factors on the ensembles with $a^{-1}\approx 3.61~\mathrm{GeV}$ using $m_\pi\approx 500~\mathrm{MeV}$ and $m_h=m_c$ (red), $m_h=1.25^2\times m_c$ (blue) and $m_h=1.25^4\times m_c$ (green). The points show the data that were used in the global fit with those parameters: red circles for $m_h=m_c$, blue squares for $m_h=1.25^2\times m_c$ and green stars for $m_h=1.25^4\times m_c$ (green). The solid black line gives the result at $m_b=4.528\times m_c$.}
\label{fig:mh_dependence}
\end{figure}

A key issue in this work is the size of the dependence on the heavy quark mass, $m_h$. Figure~\ref{fig:mh_dependence} shows how the form factors vary with this value on the ensemble with $a^{-1}\approx 3.61~\mathrm{GeV}$ and with $m_\pi\approx 500~\mathrm{GeV}$. Dashed line (point) colours indicate the value of the heavy quark mass: red (circles) for $m_h=m_c$, blue (squares) for $m_h=1.25^2\times m_c$ and green (stars) for $m_h=1.25^4\times m_c$. The solid black line indicates an extrapolation to the physical value, which we have taken as $m_b=4.528\times m_c$ from~\cite{Maezawa:2016vgv} and~\cite{Chakraborty:2014aca}. Again we find that the $f_1(v \cdot k_\pi)+f_2(v \cdot k_\pi)$ has a larger dependence than $f_2(v\cdot k_\pi)$, but that it is not severe.

\subsection{Form Factors}
\begin{figure}[!bt]
  \centerline{
    \includegraphics[width=0.85\textwidth]{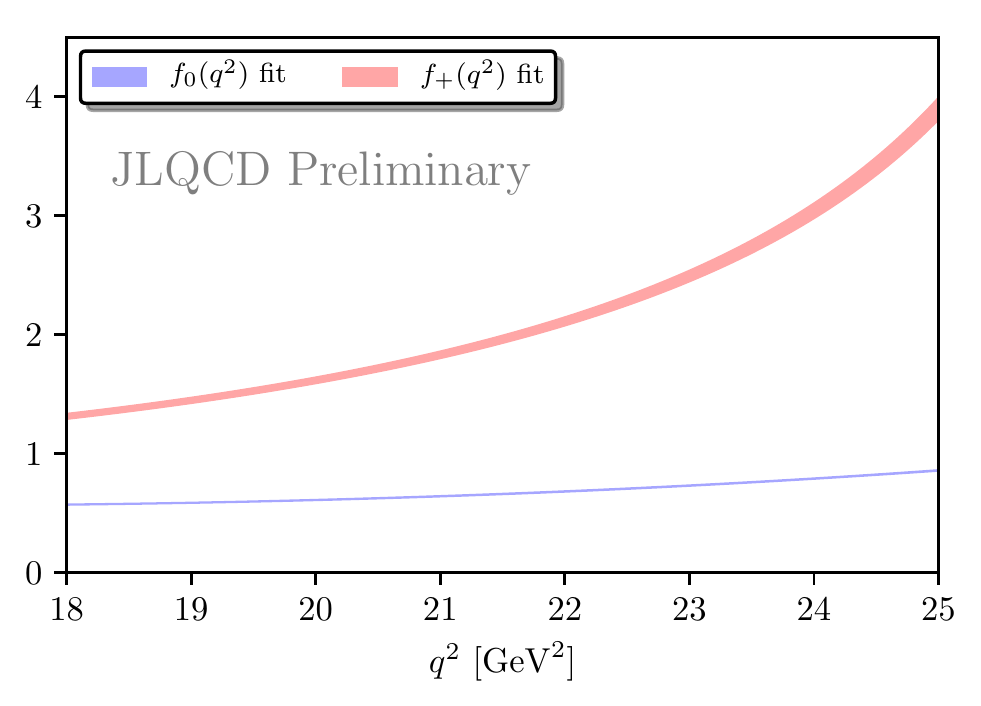}
    }
\caption{Bands denoting form factors $f_+(q^2)$ (top) and $f_0(q^2)$ (bottom) derived by using~(\ref{eq:form_factor_transform}) to convert the form factors $f_1(v\cdot k_\pi)+f_2(v \cdot k_\pi)$ and $f_2(v \cdot k_\pi)$ obtained through our global fits. The widths of the bands indicate the statistical error only.}
\label{fig:form_factors}
\end{figure}

Using the global fit results from the form factors $f_1(v\cdot k_\pi)+f_2(v \cdot k_\pi)$ and $f_2(v\cdot k_\pi)$ we can convert back to the initial parametrisation through~(\ref{eq:form_factor_transform}). We show the resulting values of the scalar form factor, $f_0(q^2)$ (bottom band), and the vector form factor, $f_+(q^2)$ (top band), in Figure~\ref{fig:form_factors}. This latter form factor is the one that will be of relevance in determining the value of the CKM matrix element $\left\vert V_{ub} \right\vert$, as shown in~(\ref{eq:ckm_extraction}).

The width of the bands indicate the central values plus and minus a standard deviation, where the errors are entirely statistical. We restrict the $q^2$ values to the region $(M_B-E_\pi)^2$ using the \textit{physical} $B$ meson mass, and the range of pion energies we obtained on the lattice.

\section{Ongoing Work}
We continue to gather statistics for the $B\rightarrow \pi \ell \nu$ decay. Using a larger volume ($48^3\times 96$) with inverse lattice spacing $a^{-1}\approx 2.45~\mathrm{GeV}$, we are also generating correlators with $m_\pi=230~\mathrm{MeV}$.

Calculating the form factors at smaller values of $q^2$ requires a more sophisticated approach than the fit we presented here. We intend to extend our reach by performing a fit to a $z$-parametrisation.  The result for $f_+(q^2)$ obtained in this way can be used to extract a value for the CKM matrix element $|V_{ub}|$ by fitting with the $B\rightarrow \pi \ell \nu$ differential decay rate obtained from experimental data.

\section*{Acknowledgements}
We thank members of the JLQCD collaboration for discussions and for providing the computational framework. Numerical calculations were performed on the Blue Gene/Q supercomputer at KEK under its Large Scale Simulation Program (No. 16/17-14). This work is supported in part by JSPS KAKENHI Grant Number JP26247043 and by the Post-K supercomputer project through the Joint Institute for Computational Fundamental Science (JICFuS).

\bibliography{sources}

%%%%%%%%%%%%%%%%%%%%%%%%%%%%%%%%%%%%%%%%%%%%%%%%%%%%%%%%%%%%%%%%%%%%%%%%%%%%%
\end{document}